\documentclass[usenatbib]{article}
\usepackage{times,graphicx,natbib,a4wide}

\begin{document}
	
\title{Spatio-temporal Constraints on the Zoo Hypothesis, and the Breakdown of Total Hegemony}
\author{Duncan H. Forgan$^1$}
\maketitle

\noindent $^1$Scottish Universities Physics Alliance (SUPA), Institute for Astronomy, University of Edinburgh, Blackford Hill, Edinburgh, EH9 3HJ, UK \\

\noindent \textbf{Word Count: 3,190} \\
\noindent \textbf{Email:} dhf@roe.ac.uk \\

%\newpage

\begin{abstract}

\noindent The Zoo Hypothesis posits that we have not detected extraterrestrial intelligences (ETIs) because they deliberately prevent us from detecting them.  While a valid solution to Fermi's Paradox, it is not particularly amenable to rigorous scientific analysis, as it implicitly assumes a great deal about the sociological structure of a plurality of civilisations.  Any attempt to assess its worth must begin with its most basic assumption - that ETIs share a uniformity of motive in shielding Earth from extraterrestrial contact. This motive is often presumed to be generated by the influence of the first civilisation to arrive in the Galaxy.  I show that recent work on inter-arrival time analysis, while necessary, is insufficient to assess the validity of the Zoo Hypothesis (and its related variants).  The finite speed of light prevents an early civilisation from exerting immediate cultural influence over a later civilisation if they are sufficiently distant.  I show that if civilisation arrival times and spatial locations are completely uncorrelated, this strictly prevents the establishment of total hegemony throughout the Galaxy.  I finish by presenting similar results derived from more realistic Monte Carlo Realisation simulations (where arrival time and spatial locations are partially correlated).  These also show that total hegemony is typically broken, even when the total population of civilisations remains low.  The Zoo Hypothesis is therefore only justifiable on weak anthropic grounds, as it demands total hegemony established by a long-lived early civilisation, which is a low probability event.  In the terminology of previous studies of solutions to Fermi's Paradox, this confirms the Zoo Hypothesis as a "soft" solution.  However, an important question to be resolved by future work is the extent to which many separate hegemonies are established, and to what extent this affects the Zoo Hypothesis.

\end{abstract}

\section{Introduction}

\noindent Fermi's Paradox remains one of the most important questions in astrobiology and SETI (the Search for Extraterrestrial Intelligence).  In short, the Paradox asks: ``If the timescale for a civilisation forming and crossing the Galaxy is much lower than the age of the Galaxy, why have we seen no evidence of extraterrestrial intelligence?''  A variety of solutions have been formulated, ranging from

\begin{enumerate}
 \item There are very few or no other civilisations (The Rare Earth Hypothesis, \citealt{rare_Earth})
\item They are here, but in secret (sometimes jokingly referred to as ``the X-Files Hypothesis'')
\item Civilisations find it too expensive or damaging to colonise the Galaxy, and instead choose not to (The Sustainability Solution, e.g. \citealt{empire, Haqq2009})
\end{enumerate}

\noindent and so on.  Thorough reviews of the Paradox can be found in \citet{BrinG.D.1983} and \citet{fermi_review}.  This paper shall focus on the so-called Zoo Hypothesis or Zoo Solution - the concept that Mankind is being deliberately shielded from (or ignored by) the Galactic community, either because humans are deemed ``too primitive'', interfering with civilisations at our stage of development is dangerous, or simply because we provide an interesting example of a developing civilisation for alien anthropologists or historians \citep{Ball1973}.  We should expect that such a policy can only be enforced in two circumstances:

\begin{enumerate}
 \item Only one extraterrestrial civilisation exists within contact range
 \item A universal legal policy or treaty exists which forbids signatories to interfere (e.g. \citealt{Freitas1977}'s description of potentially global Earth Law)
\end{enumerate}

\noindent \citet{Hair2011} has studied the second circumstance in more detail.  If such a treaty is to exist, a uniformity of motive is required amongst the Galactic community, something which grows less and less likely as more civilisations arise.  After all, only one civilisation must break the treaty to interfere with Earth's development, making the Zoo Hypothesis a ``soft'' solution (using the definitions in \citealt{fermi_review} and references within).  To maintain uniformity of motive, all civilisations must evolve towards similar standards and conduct, an eventuality greatly assisted if there is a dominant culture already in existence.  Hair proposes that the first Galactic civilisation could provide this dominant culture.  This first civilisation can be present at the beginning of other civilisations, shepherding them towards a common culture and the required uniformity of motive.

Hair studies the distribution of the first Inter-Arrival Time (referred to hereafter as $IAT_1$):

\begin{equation} IAT_1 = t_2 - t_1 \end{equation}

\noindent Where $t_i$ is the time at which civilisation $i$ becomes ``intelligent'' (i.e. it constructs its first radio telescope or equivalent, and can receive interstellar messages).  If $t_i$ is distributed as a Gaussian, it can be demonstrated that $IAT_1$ is representable by an inverse exponential distribution \citep{Snyder1991}.  In this simplified model, Hair shows that the first civilisation may exist for up to 300 Myr alone in the Galaxy.  This may allow the ``chain of culture'' to be set up, but how likely is the chain to break?  For the Zoo Hypothesis to be strong, we require that this hegemony must be passed from civilisation to civilisation without the chain being broken.  

Once a civilisation arises, the maximum speed at which it can spread its influence is the speed of light \emph{in vacuo}.  A time limit is therefore imposed - if $IAT_{i+1}$ between civilisations $i$ and $i+1$ is short compared to their spatial separation, then we expect the chain of culture to be broken - civilisation $i+1$ will arise before civilisation $i$ can communicate its cultural identity, and uniformity of motive is likely to fail.  

It is true that hegemony may continue if the first civilisation is sufficiently long-lived that $t_3-t_1$ is large compared to the relative separation of civilisations 1 and 3, and so on up to higher civilisation numbers.  As we will see later in this paper, total hegemony will require the first civilisation (or at least one early civilisation) to persist for an extremely long time, further softening this variant of the Zoo Hypothesis, as we cannot say anything conclusive currently about the lifetime of civilisations.  In this work, I will focus on the harder alternative - that the chain of culture is not forced by any individual civilisation, but instead by all civilisations in their turn. The chain of culture is therefore only as strong as its weakest link. 

While inter-arrival time analysis is an important and necessary component of any study of the Zoo Hypothesis, we cannot forget the constraints imposed by the spatial distribution of civilisations \citep{Freitas1980}.  This paper will investigate combined spatio-temporal constraints in two separate studies. The first will consider a simple toy model, much in the same fashion as \cite{Hair2011}, with the added constraints of spatial distribution.  In this model, the arrival times and spatial locations of each civilisation are strictly uncorrelated.

The second study will use more realistic datasets, culled from Monte Carlo Realisation (MCR) simulations of life and intelligence in the Milky Way \citep{mcseti1,mcseti2}.  Because of its explicit incorporation of the Milky Way's Star Formation History, Age-Metallicity Relation and metallicity gradients, the arrival times and spatial locations of each civilisation will be somewhat correlated.  

The results of both studies indicate that the added constraints imposed by spatial location will work to break the hegemony established by the first civilisation, and that the chain of culture is in fact several separate chains of culture.  While this does not completely eradicate the possibility of the Zoo Hypothesis being correct, it seriously impairs it.  If one partial hegemony advocates interference and begins doing so, it weakens the Zoo Solution in the same fashion as single civilisations do when they make a similar choice.  

The paper is set out as follows: I describe the construction of the first (toy) model in section \ref{sec:toy} and the MCR model in section \ref{sec:MCR}; I describe how hegemony is measured in both models in section \ref{sec:heg}; I discuss the results of both models in section \ref{sec:results} and I summarise the work in section \ref{sec:conclusions}.

\section{The Simple Toy Model}\label{sec:toy}

\noindent In the same vein as \citet{Hair2011}, the simple toy model proposes that the distribution of the civilisation arrival times $t_i$ be Gaussian-distributed.  We will see in the following sections that this is a reasonable approximation to that found from the MCR model.  For the spatial location of the civilisations, I assume that they are evenly distributed in an annulus, with inner radius 7 kpc and outer radius 9 kpc ($1$ kpc $= 1000$ pc $\approx 3000$ light years).  This corresponds roughly to the location of the Galactic Habitable Zone \citep{GHZ}, which describes the region of the Galaxy most amenable to the development of life and intelligence (our Sun sits near the middle of this annulus at around 8 kpc).  I assume for simplicity that the annulus has zero thickness - this will not affect the overall result significantly.

Having sampled $t_i$ from a Gaussian, the spatial location of $i$, given in cylindrical co-ordinates $(r_i, \phi_i)$, is sampled as described above.  The parameters of civilisation $i$ are randomly sampled without relation to civilisation $i+1$, ensuring such civilisations are uncorrelated.  Also, there is no relationship between $t_i$ and the location of civilisation $i$, so the set of parameters for any individual civilisation is also uncorrelated.

Each model was run for 10,000 iterations.   A total of five models were run, for varying values of the mean and standard deviation of the Gaussian distribution for $t_i$ ($\mu_t$ and $\sigma_t$ respectively).  These correspond to the same models run by \citet{Hair2011}, who uses 100 iterations in his work: we select 10,000 to confirm these results at higher precision, providing a strong framework for investigating any subsequent spatio-temporal constraints.

\section{The MCR Model}\label{sec:MCR}

\noindent The MCR model used in this work was first described in \citet{mcseti1}.  The data used in this analysis was first presented in \citet{mcseti2} - the reader is referred to these works for a detailed description of the model, but for convenience I shall present a brief summary here.

The model generates a Galaxy of $N_*$ stars, whose properties (e.g. mass, location in the Galaxy, age, luminosity, metallicity) are sampled according to the known statistical distributions for these parameters - the Initial Mass Function \citep{IMF}, the Star Formation History \citep{Rocha_Pinto_SFH}, the Age-Metallicity Relation \citep{Rocha_Pinto_AMR}, and the structure of the stellar component of the Galaxy \citep{Ostlie_and_Caroll}, including its metallicity gradients \citep{z_grad}.  Having created these stars, some will be assigned planets, based on their metallicity \citep{Wyatt_z}.  The properties of these planets are again sampled according to their expected statistical distributions.  In this instance, theoretical distributions are preferred to empirical data, as using current exoplanet observations introduces significant bias \citep{mcseti1}.

Having generated a Galaxy containing billions of stars and planets, a hypothesis for life is imposed on the model (for example, ``if a planet resides within the Stellar Habitable Zone, then life can exist upon it'').  This gives a population of inhabited planets, and the evolution of these biospheres is calculated using stochastic equations (see \citealt{mcseti1} for details).  This produces one realisation of the civilisation population.  In total, 30 separate realisations of the model were generated, to allow averaging of the output variables and characterisation of the random uncertainty.

\begin{figure}
 \begin{center}
 \includegraphics[scale=0.6]{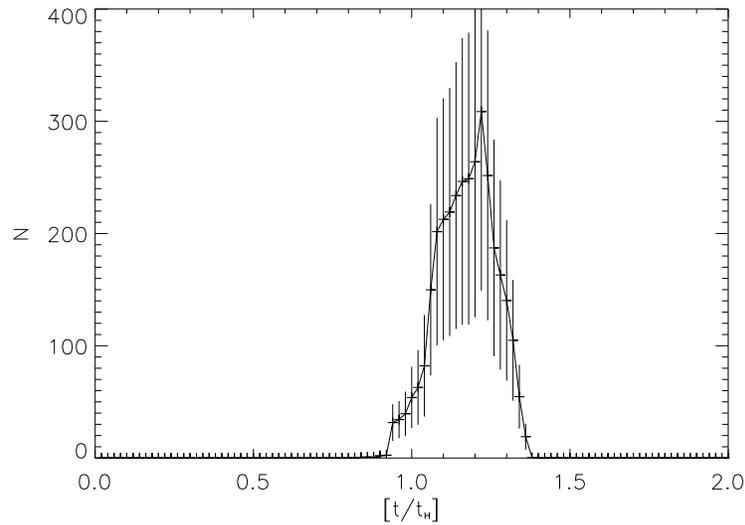} \\
 \caption{The number of existing civilisations in the ``Rare Earth'' MCR model as a function of time, in units of the Hubble Time ($t_H = 13,700$ Myr).  The mean from 30 iterations of the model is displayed, with error bars indicating one sample standard deviation.\label{fig:signal_rare}}
 \end{center}
 \end{figure}

\begin{figure}
 \begin{center}
 \includegraphics[scale=0.6]{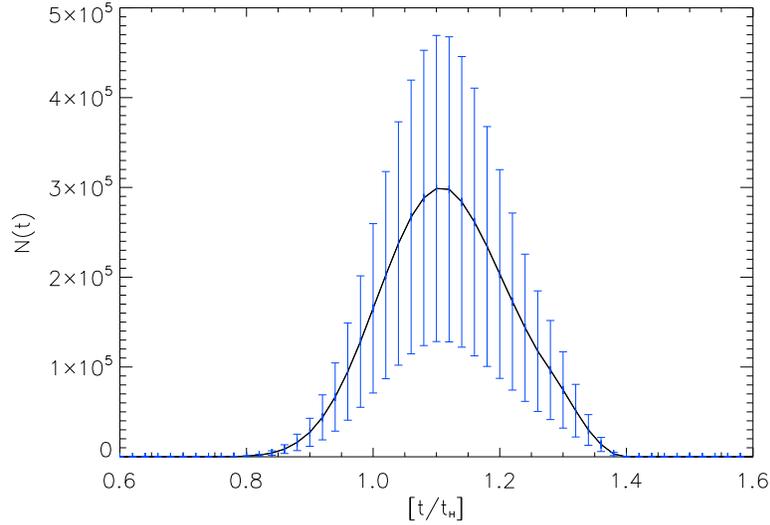} \\
 \caption{The number of existing civilisations in the ``Baseline'' MCR model as a function of time, in units of the Hubble Time ($t_H = 13,700$ Myr).   The mean from 30 iterations of the model is displayed, with error bars indicating one sample standard deviation.\label{fig:signal_baseline}}
 \end{center}
 \end{figure}

The key data produced by the MCR model (for our purposes) are the multiplicity of civilisations (Figures \ref{fig:signal_rare} \& \ref{fig:signal_baseline}), their arrival times and their spatial distribution.  Through the combined action of the Age-Metallicity relation and the Galaxy's metallicity gradient, a weak correlation is obtained between the civilisation's arrival time and its cylindrical distance from the Galactic Centre, $r$.  However, the other cylindrical co-ordinates ($\phi$ and $z$ respectively) have no such correlation.  The three-dimensional nature of these simulations (in comparison to the simple toy model's two dimensions) will allow civilisation spacing to be even larger - at this point it is unclear whether this effect will supersede that of weak spatio-temporal correlation.

Two separate data sets will be investigated for this model.  The first is a ``Rare Earth'' type simulation, where the conditions for complex life to develop are stringent (see \citealt{mcseti2} for details), and a ``baseline'' simulation where the conditions for complex life are significantly less stringent.  As such, the ``Rare Earth'' simulation has a significantly lower population than the baseline, allowing us to test the ``crowded galaxy'' concepts of \citet{Hair2011}.

\section{Measuring Hegemony in Both Models}\label{sec:heg}

\noindent We must place spatial and temporal constraints on the same footing, and construct a parameter that shows definitively when hegemony can be established, and when it must be broken.  We can construct such a quantity, $H$, by multiplying the inter-arrival time by the speed of light, $c$:

\begin{equation} H_{i} = c^2\left(IAT_{i}\right)^2 - \left|\mathbf{r}_{i+1} - \mathbf{r}_{i}\right|^2 \label{eq:H}\end{equation}

\noindent where $\mathbf{r}_i$ is the vector describing the location of civilisation $i$ (relative to the Galactic Centre). This is similar to the construction of the space-time interval in Minkowskian space-time, and has the following limits:

\begin{equation} H_{i}  \left\{
\begin{array}{l l }
< 0 & \quad \mbox{Civilisation $i$ cannot send a signal to $i+1$ before $i+1$ becomes intelligent} \\
 = 0 & \quad \mbox{Civilisation $i$'s signal arrives at $i+1$ at the same instant $i+1$ becomes intelligent} \\
> 0 & \quad \mbox{Civilisation $i$ can send a signal to $i+1$ before $i+1$ becomes intelligent} \\
\end{array} \right. \end{equation}

\noindent This will show unequivocally where the ``chain of culture'' is, and is not, broken (cf the connectivity parameters described in \citealt{mcseti2} and \citealt{SKA}).  It also allows us to quantify the strength of the influence civilisation $i$ holds over civilisation $i+1$.  If $H_{i}=x$ (in these units, where distance is in kpc, time in Myr), then we can say that civilisation $i$ is capable of broadcasting signals to $i+1$ for $\sqrt{x}/c$ Myr before civilisation $i+1$ becomes sufficiently intelligent to receive such signals.  It is straightforward to create a variant of $H$ which determines whether civilisation $i$ can be physically present at $i+1$'s homeworld before they become intelligent:

\begin{equation} \tilde{H}_i = v_{craft,i}^2\left(IAT_{i+1}\right)^2 - \left|\mathbf{r}_{i+1} - \mathbf{r}_{i}\right|^2 \label{eq:tildeH}\end{equation}

\noindent Where $v_{craft,i}$ is the maximum velocity of spacecraft belonging to civilisation $i$.

\section{Results and Discussion}\label{sec:results}

\subsection{The Toy Model}

\begin{figure}
 \begin{center}
 \includegraphics[scale=0.6]{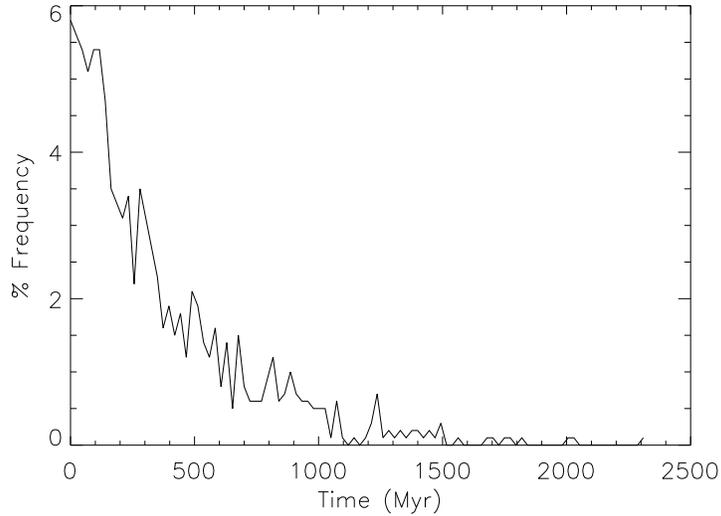} \\
 \caption{The distribution of $IAT_1$, from 10,000 iterations of toy model 3 (where $\mu_t = 13000$ Myr, $\sigma_t = 1500$ Myr, see Table \ref{tab:params}). \label{fig:binned_IAT1}}
 \end{center}
 \end{figure}

\noindent The toy model was run using the parameters of \citet{Hair2011}, which can be seen in Table \ref{tab:params}.  I recover the same distribution of $IAT_1$, as shown in Figure \ref{fig:binned_IAT1}.  The maximum $IAT_1$ obtained in this set (from 10000 iterations) is approximately 2 Gyr, seemingly a long period of time for the first civilisation to establish itself.  

\begin{table}
\centering

  \caption{Summary of the toy model parameters investigated in this work.  All models were subjected to 10,000 iterations.\label{tab:params}}
  \begin{tabular}{c ||ccc}
  \hline
  \hline
   Model & $N_{civ}$ & $\mu_t$ (Myr) &  $\sigma_{t}$ (Myr)  \\  
 \hline
   1 & 100	& 13,000 & 1500 \\
   2 & 1000	& 13,000 & 1500 \\
   3 & 10000	& 13,000 & 1500 \\
   4 & 10000	& 13,000 & 1000 \\
   5 & 10000	& 13,000 & 650 \\
 \hline
  \hline
\end{tabular}
\end{table}

\begin{figure}
 \begin{center}
 \includegraphics[scale=0.6]{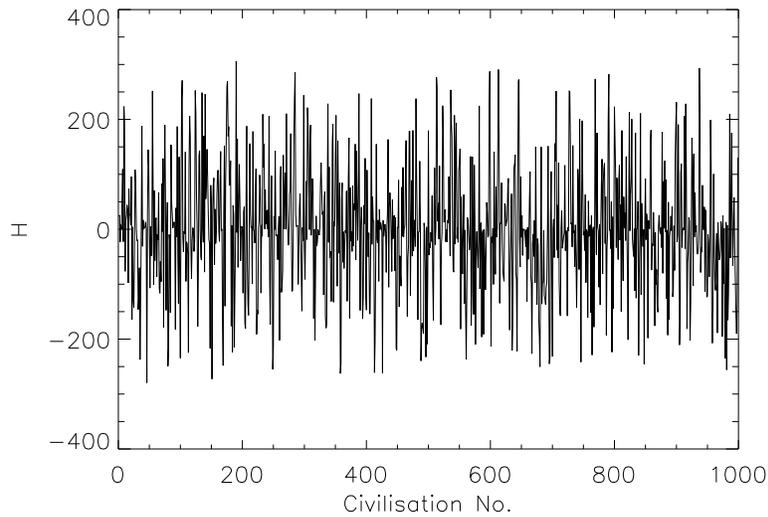} \\
 \caption{The breakdown of total hegemony.  The graph displays the hegemony parameter for 1000 civilisations in one iteration of the model using parameter set 3 (where $\mu_t = 13000$ Myr, $\sigma_t = 1500$ Myr, see Table \ref{tab:params}).  Areas where the hegemony parameter are negative shows the spatial separation to be too great for one civilisation to communicate its culture to the next, showing that total hegemony does not occur. This behaviour is found across all 10,000 civilisations - only the first 1,000 are shown for the sake of clarity. \label{fig:heg_breakdown}}
 \end{center}
 \end{figure}

However, the hegemony parameter $H$ tells a different story.  Figure \ref{fig:heg_breakdown} shows the value of $H$ for each civilisation in one iteration, using parameter set 3.  $H$ oscillates rapidly between positive and negative values, showing that the chain of culture is repeatedly broken.  The amplitude of $H$ is roughly the same both at positive and negative values, and appears to be centred around zero, an intuitive result for uncorrelated variables.  We can confirm this result by calculating the mean value of $H$ for all the civilisations in each iteration.  
 
\begin{figure}
 \begin{center}
 \includegraphics[scale=0.6]{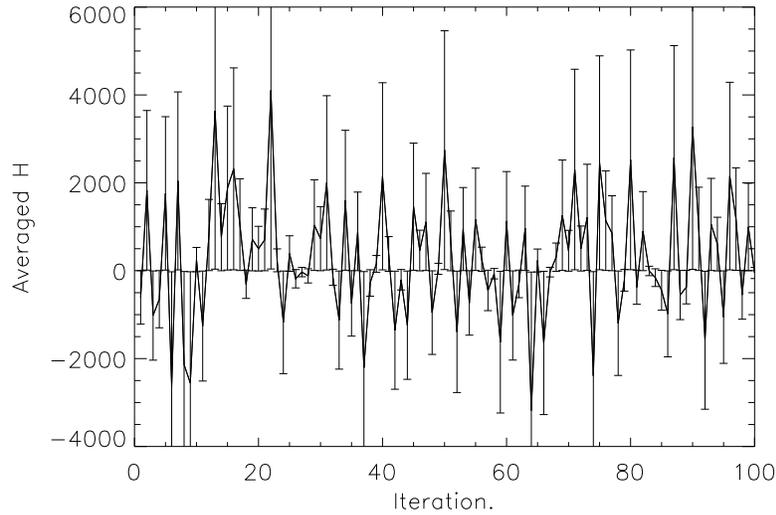} \\
 \caption{The mean value of the hegemony parameter, averaged over 1000 civilisations, from 100 iterations of model 3 (where $\mu_t = 13000$ Myr, $\sigma_t = 1500$ Myr).  Also plotted are the sample standard deviations calculated over the 1000 civilisations, showing that each value of the hegemony parameter is approximately consistent with zero at $1 \sigma$ (and strictly consistent with zero at $3 \sigma$).  The same behaviour is found for all 10,000 iterations of model 3 - again, only the first 100 are shown for the sake of clarity. \label{fig:mean_heg}}
 \end{center}
 \end{figure}

Figure \ref{fig:mean_heg} shows while the mean value of $H$ (averaged over all civilisations), $\bar{H}$, varies between positive and negative values, the standard deviation of $H$, $\sigma_H \approx \bar{H}$.  Therefore, $\bar{H}$ is approximately consistent with zero to within $1\sigma$, and less than zero to within $3\sigma$.  This behaviour is found in all five sets of model parameters, and is clearly a direct consequence of decoupling the spatial and temporal co-ordinates of civilisations.  This underlines the importance of considering both space and time together in such analyses.  We can also be certain that any variant $\tilde{H}$ based on colonisation speed (equation (\ref{eq:tildeH})) will most likely be consistent with zero, if not a negative value, as $c$ is the ultimate speed limit in the Universe.

What this analysis does not capture is the establishment of partial hegemonies, which are likely to be localised in space but will not necessarily constitute an unbroken consecutive chain of civilisations.  A more detailed study of the growth of individual groups is left for further work, but what can be stated is that if more hegemonies of civilisations exist in one Galaxy, less uniformity of motive exists (much as it does for unconnected individual civilisations).  The hegemony is simply redefined as the basic unit of civilisation, which now includes many biospheres as opposed to one.

\subsection{The MCR Model}

\subsubsection{The Rare Earth Model}

 \begin{figure}
 \begin{center}
 \includegraphics[scale=0.6]{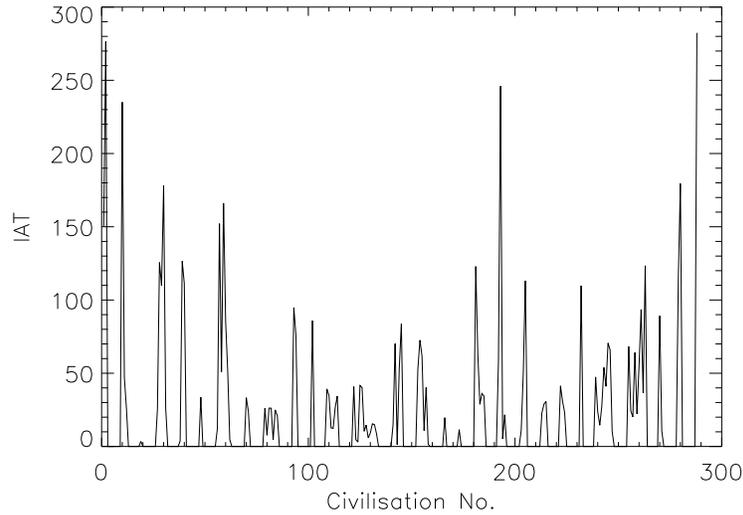} \\
 \caption{The inter-arrival time for each civilisation, as a function of civilisation number for one iteration of the ``Rare Earth'' simulation.  The maximum IAT is approximately 300 Myr, with a minimum of zero.  The minimum is  zero because some star systems contain more than one inhabited planet due to colonisation, and hence the inter-arrival time is effectively zero. \label{fig:IAT_rare}}
 \end{center}
 \end{figure}

\begin{figure}
 \begin{center}
 \includegraphics[scale=0.6]{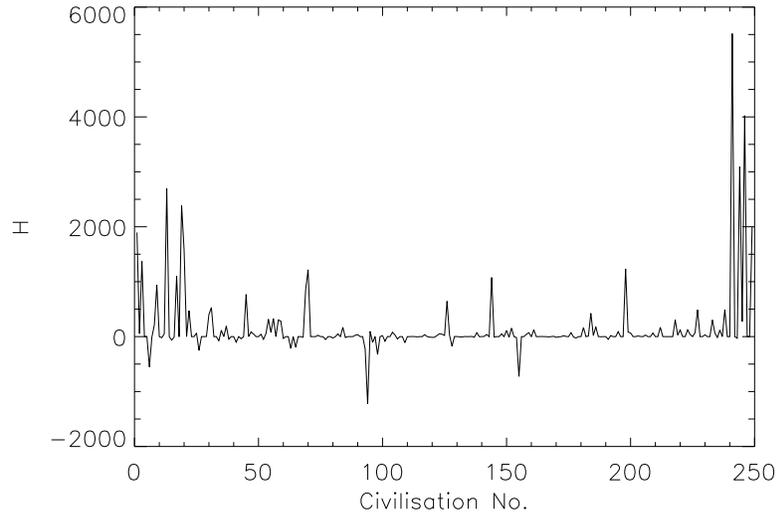} \\
 \caption{The hegemony factor for each civilisation, as a function of civilisation number for one iteration of the ``Rare Earth'' simulation.  Again, $H$ becomes zero frequently due to civilisations colonising multiple planets in their home star system.  While $H$ is often positive, the chain is broken several times.  \label{fig:heg_rare}}
 \end{center}
 \end{figure}

\noindent Does partially correlating spatial and temporal position improve the establishment of hegemony? The evidence would suggest otherwise - Figure \ref{fig:heg_rare} shows $H$ against civilisation number for several iterations of the ``Rare Earth'' model.  Civilisations can colonise multiple planets in the same star system, hence there are chains of $H=0$ scattered throughout the simulation, but there is always one value of $H<0$ in the chain for all 30 iterations, confirming that the chain of culture is broken, even in a sparsely populated Galaxy with inter-arrival times peaking at around 300 Myr.  This suggests that total hegemony is unlikely even in relatively empty Galaxies (although the resulting hegemony could still be quite large).  Note in the early phases that $H \sim 2000$, i.e. these civilisations can send signals to the next civilisation for approximately 50 Myr before the next civilisation receives them.  These early, smaller hegemonies are therefore expected to be well-established (although such hegemonies occur at different civilisation numbers depending on the iteration of the simulation, so any discussion of when strong hegemonies arise is beyond the bounds of this work).  Also, while $H$ may be large, $\tilde{H}$ is likely to be much smaller.  Mankind's current velocity record is around $10^{-5} c$ - as a lower limit on colonisation speeds, this would make $\tilde{H}$ extremely small and most likely negative.  If physical presence is required to initiate hegemonies, it is essentially impossible in the ``Rare Earth'' model.

\subsubsection{The Baseline Model}

\begin{figure}
 \begin{center}
 \includegraphics[scale=0.6]{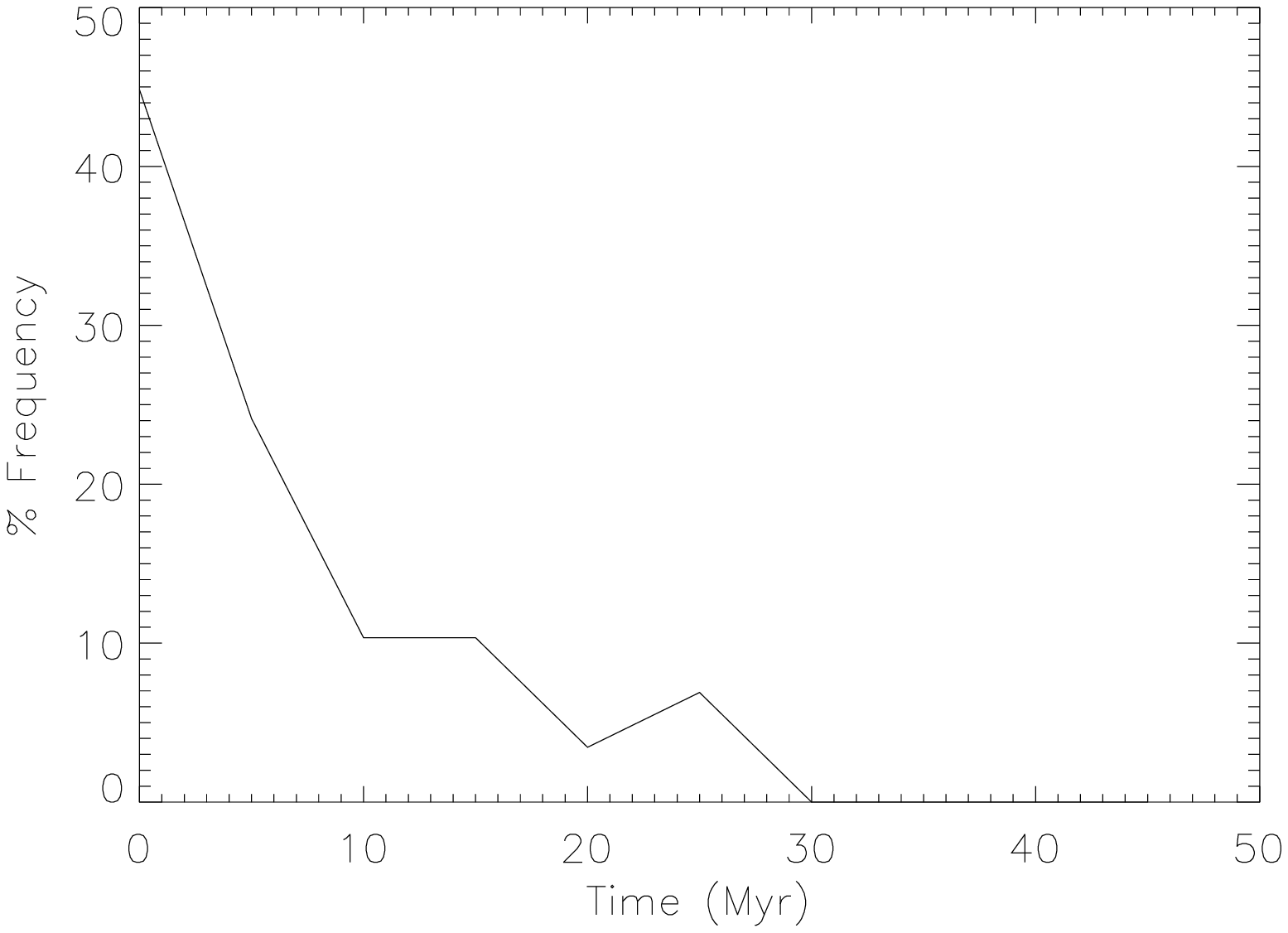} \\
 \caption{Binned values of $IAT_1$ for 30 iterations of the ``Baseline'' simulation. These statistics exclude inter-arrival times calculated between inhabited planets in the same star system. \label{fig:IAT_binned_baseline}}
 \end{center}
 \end{figure}

\begin{figure}
 \begin{center}
 \includegraphics[scale=0.6]{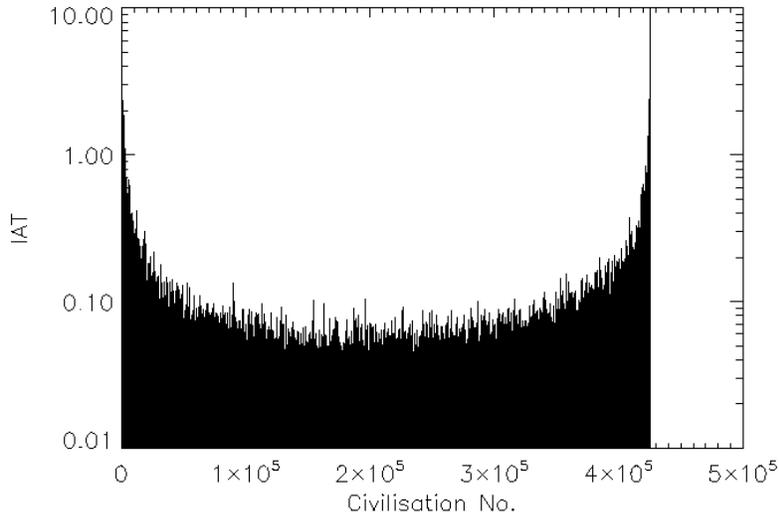} \\
 \caption{The inter-arrival time for each civilisation, as a function of civilisation number for one iteration of the ``Baseline'' simulation.  The maximum IAT is approximately 1 Myr, with a minimum of 0.  The minimum is exactly zero because some star systems contain more than one civilisation, and hence the inter-arrival time is effectively zero. \label{fig:IAT_baseline}}
 \end{center}
 \end{figure}

\noindent In the second, baseline model, there are significantly more civilisations, arriving within a similar timescale (cf Figures \ref{fig:signal_baseline}), presenting a more crowded galaxy.  This increase in numbers, within approximately the same timescale for arriving (given by the width of the Gaussian in Figure \ref{fig:signal_baseline}), reduces the inter-arrival times significantly.  Figure \ref{fig:IAT_binned_baseline} shows that 60\% of civilisations have inter-arrival times less than 10 Myr.  This is confirmed by Figure \ref{fig:IAT_baseline}, which shows again that the typical IAT is less than 1 Myr.  But will this affect $H$, i.e. will this decreased IAT be accompanied by a sufficiently decreased spatial separation?  As the civilisations tend to occupy the same regions of the synthetic Galaxy (i.e. the Galactic Habitable Zone), it might be reasonable to assume that $H$ might be maintained at values similar to that shown for the Rare Earth model.  This however is not the case - in fact $H$ is even more negative (with absolute values increased by as much as an order of magnitude).  Note also that there are essentially no circumstances where $H$ is significantly larger than zero.  Again, if hegemony is limited by some colonisation speed, this proves without doubt that this Baseline Model would never develop strong hegemonies.

\begin{figure}
 \begin{center}
 \includegraphics[scale=0.6]{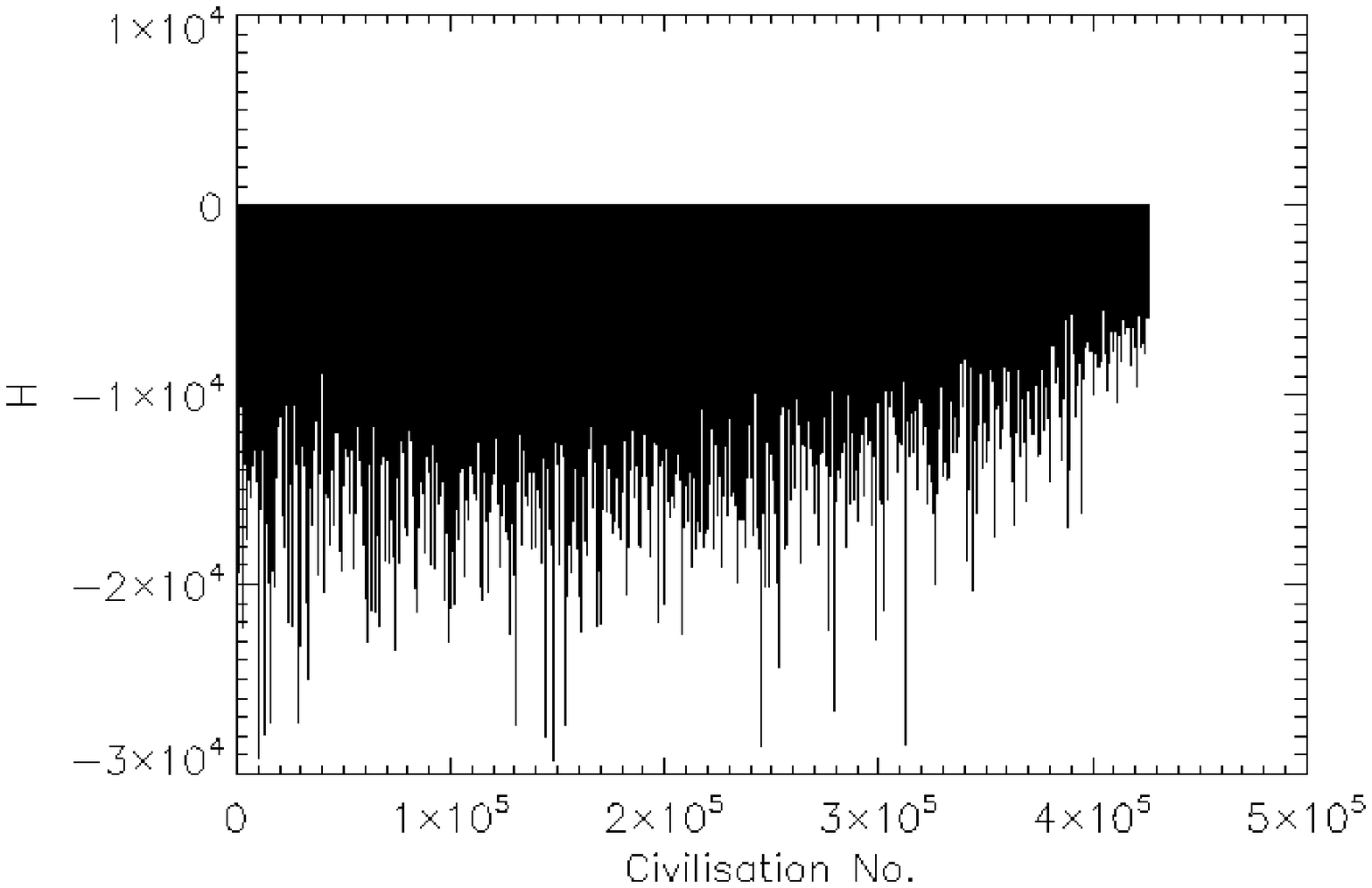} \\
 \caption{The hegemony factor for each civilisation, as a function of civilisation number for one iteration of the ``Baseline'' simulation.  $H$ is typically less than zero across the entire history of this Galaxy - the chain of culture is never truly established.  \label{fig:heg_baseline}}
 \end{center}
 \end{figure}

\section{Conclusions}\label{sec:conclusions}

\noindent I have investigated a key assumption of the Zoo Solution to Fermi's Paradox - generally that many civilisations establish a uniformity of motive, and more particularly that this motive emanates from the first Galactic civilisation, and remains in an unbroken ``chain of culture'' across Galactic history.  I have performed a combined spatio-temporal inter-arrival analysis for civilisations in the Galaxy, using both a simple toy model and more realistic data obtained from Monte Carlo Realisation (MCR) simulations.  

In the toy model, spatial and temporal location of civilisations are uncorrelated, and the establishment of a constant ``chain of culture'' is broken quickly and easily.  In fact, on average the hegemony parameter $H$, which measures the space-time separation of the arrival of two civilisations, is consistent with negative values to 3$\sigma$, implying that hegemony is never truly established in these circumstances.  This result remains constant despite varying the parameters of the inter-arrival time distribution, and appears to rule out total hegemony by any one culture.  

These trends are also found in the data from MCR simulations, despite spatial and temporal location being weakly correlated.  Total hegemony is broken both in the case where there are few civilisations and in the case where the Galaxy is crowded.  What is more likely is that partial hegemonies are established by several civilisations - the total number of ``empires'' in this case is unclear, and requires further analysis.  We must not disregard either the possibility that empires are not pursued by extraterrestrial intelligences, due to their being economically untenable \citep{empire, Haqq2009}.  This aside, the more hegemonies exist, the less likely it is that uniformity of motive can be established (in the exact same manner as if unconnected civilisations attempted to establish uniformity without contact or cultural exchange).  In this revised model, hegemonies are the basic unit of civilisation, where a ``civilisation'' incorporates many different sub-civilisations (much as the Roman Empire encompassed other civilisations such as the Greeks and the Britons).

The only exception to these conclusions is if any one civilisation survives for a significantly long time to ``repair'' the chain of culture wherever it is broken.  Postulating such a civilisation weakens the Zoo Hypothesis by requiring its occurence early on in the chain, which reduces its probability significantly.  Of course, the large number of galaxies in the Universe will increase the absolute occurrence of such low-probability events, allowing the required special circumstances for the ``Eerie Silence'' to exist in some galaxies \citep{Davies2010}.  In this sense, the Zoo Solution is only defensible using the weak anthropic principle (glibly, "things are the way they are because we observe them").  Therefore, we cannot invalidate the Zoo Hypothesis, but it is definitely a "soft" solution, and in that respect unsatisfactory as a means of resolving Fermi's Paradox.

\section{Acknowledgements}

\noindent This work has made use of the resources provided by the Edinburgh Compute and Data Facility (ECDF, http://www.ecdf.ed.ac.uk/). The ECDF is partially supported by the eDIKT initiative (http://www.edikt.org.uk).

\bibliographystyle{mn2e} % (must include a bibliography style)
\bibliography{IAT1_1}

\begin{thebibliography}{21}
\expandafter\ifx\csname natexlab\endcsname\relax\def\natexlab#1{#1}\fi

\bibitem[{Ball(1973)}]{Ball1973}
Ball J., 1973, Icarus, 19, 347

\bibitem[{Brin(1983)}]{BrinG.D.1983}
Brin G.~D., 1983, QJRAS, 24, 283

\bibitem[{Cirkovic(2009)}]{fermi_review}
Cirkovic M., 2009, Serbian Astronomical Journal, 178, 1

\bibitem[{Cirkovic(2008)}]{empire}
Cirkovic M.~M., 2008, J. Br. Interplanet. Soc., 61, 246

\bibitem[{Davies(2010)}]{Davies2010}
Davies P., 2010, {The Eerie Silence: Are We Alone in the Universe?} Allen Lane

\bibitem[{Forgan \& Nichol(2010)}]{SKA}
Forgan D., Nichol R., 2010, International Journal of Astrobiology, 10, 77

\bibitem[{Forgan(2009)}]{mcseti1}
Forgan D.~H., 2009, International Journal of Astrobiology, 8, 121

\bibitem[{Forgan \& Rice(2010)}]{mcseti2}
Forgan D.~H., Rice K., 2010, International Journal of Astrobiology, 9, 73

\bibitem[{Freitas(1977)}]{Freitas1977}
Freitas R.~A., 1977, Mercury, 6

\bibitem[{Freitas(1980)}]{Freitas1980}
---, 1980, J. British Interplanet. Soc., 33

\bibitem[{Hair(2011)}]{Hair2011}
Hair T.~W., 2011, International Journal of Astrobiology, 10, 131

\bibitem[{Haqq-Misra \& Baum(2009)}]{Haqq2009}
Haqq-Misra J.~D., Baum S.~D., 2009, J. Br. Interplanet. Soc., 62, 47

\bibitem[{Lineweaver {et~al.}(2004)Lineweaver, Fenner, \& Gibson}]{GHZ}
Lineweaver C.~H., Fenner Y., Gibson B.~K., 2004, Science, 303, 59

\bibitem[{Miller \& Scalo(1979)}]{IMF}
Miller G.~E., Scalo J.~M., 1979, ApJs, 41, 513

\bibitem[{Ostlie \& Carroll(1996)}]{Ostlie_and_Caroll}
Ostlie D.~A., Carroll B.~W., 1996, {An Introduction to Modern Stellar
  Astrophysics}, Ostlie D.~A., Carroll B.~W., eds. Pearson Education

\bibitem[{Rocha-Pinto {et~al.}(2000{\natexlab{a}})Rocha-Pinto, Maciel, Scalo,
  \& Flynn}]{Rocha_Pinto_SFH}
Rocha-Pinto H.~J., Maciel W.~J., Scalo J., Flynn C., 2000{\natexlab{a}}, A\&A,
  358, 850

\bibitem[{Rocha-Pinto {et~al.}(2000{\natexlab{b}})Rocha-Pinto, Maciel, Scalo,
  \& Flynn}]{Rocha_Pinto_AMR}
---, 2000{\natexlab{b}}, A\&A, 358, 869

\bibitem[{Rolleston {et~al.}(2000)Rolleston, Smartt, Dufton, \& Ryans}]{z_grad}
Rolleston W.~R.~J., Smartt S.~J., Dufton P.~L., Ryans R.~S.~I., 2000, A\&A,
  363, 537

\bibitem[{Snyder \& Miller(1991)}]{Snyder1991}
Snyder D.~L., Miller M.~I., 1991, {Random point processes in time and space}.
  Springer-Verlag, p. 481

\bibitem[{Ward \& Brownlee(2000)}]{rare_Earth}
Ward P., Brownlee D., 2000, {Rare Earth : Why Complex Life is Uncommon in the
  Universe}, Ward P., Brownlee D., eds. Springer

\bibitem[{Wyatt {et~al.}(2007)Wyatt, Clarke, \& Greaves}]{Wyatt_z}
Wyatt M.~C., Clarke C.~J., Greaves J.~S., 2007, MNRAS, 380, 1737

\end{thebibliography}

\end{document}